\documentclass[journal=apchd5,manuscript=letter]{achemso}

\usepackage[version=3]{mhchem} % Formula subscripts using \ce{}
\usepackage[T1]{fontenc}       % Use modern font encodings
\usepackage[utf8]{inputenc}
\usepackage[english]{babel}
\usepackage{amsmath,amsfonts,amsthm,amssymb}
\usepackage{mathtools}
\usepackage{graphicx}
\usepackage{xcolor}
\usepackage[pdftex]{hyperref}
\usepackage{bm}
\usepackage[inline]{enumitem}
\usepackage{array}
\usepackage{gensymb}
\usepackage[font=footnotesize,labelfont=default]{caption}
\usepackage{verbatim}
\usepackage{mathrsfs}

\renewcommand{\eqref}[1]{Eq.~(\ref{eq:#1})}

\author{Antonio~I.~Fern{\'a}ndez-Dom{\'i}nguez}
\affiliation[]{Departamento de F{\'i}sica Te{\'o}rica de la Materia Condensada and Condensed Matter Physics Center (IFIMAC), Universidad Aut{\'o}noma de Madrid, E-28049 Madrid, Spain}
\author{Sergey~I.~Bozhevolnyi}
\affiliation[]{Center for Nano Optics, University of Southern Denmark, Campusvej 55, DK-5230~Odense~M, Denmark}
\alsoaffiliation[]{Danish Institute for Advanced Study, University of Southern Denmark, Campusvej 55, DK-5230~Odense~M, Denmark}
\author{N.~Asger~Mortensen}
\affiliation[]{Center for Nano Optics, University of Southern Denmark, Campusvej 55, DK-5230~Odense~M, Denmark}
\alsoaffiliation[]{Danish Institute for Advanced Study, University of Southern Denmark, Campusvej 55, DK-5230~Odense~M, Denmark}

\email{asger@mailaps.org}

\title {Plasmon-enhanced generation of non-classical light}

\begin{document}

\begin{abstract}
Strong light-matter interactions enabled by surface plasmons
have given rise to a wide range of photonic, optoelectronic and chemical functionalities. In recent years, the interest in this research area has focused on the quantum regime, aiming to developing
ultra-compact nanoscale instruments operating at the single (few) photon(s) level. In this perspective, we provide a general overview of recent experimental and theoretical advances as well as
near-future challenges towards the design and implementation of
plasmon-empowered quantum optical and photo-emitting devices based
on the building blocks of nanophotonics technology:
metallo-dielectric nanostructures and microscopic light sources.
\end{abstract}

\section{Introduction}
During the last decade, plasmonics has flourished as a research
field\cite{Fernandez-Dominguez:2017}, with intriguing
light-matter interactions rooted in the interplay between
electromagnetic fields and collective oscillations of free
carriers in conducting media, e.g., metals\cite{McPeak:2015,Knight:2014,Bonanni:2011,Biggins:2018} as well as doped
semiconductors\cite{Liu:2017} and 2D materials\cite{Grigorenko:2012}. Strong enhancement (several orders of magnitude) and confinement (far beyond the diffraction limit) of optical fields produced by surface plasmons (SPs) in sub-wavelength nanostructures with complex shapes and topographies\cite{Gramotnev:2014} is behind recent developments in areas as diverse as nonlinear optics\cite{Kauranen:2012}, hot-carrier catalysis\cite{Brongersma:2015,Naldoni:2017} or nanoscale colour printing\cite{Kristensen:2016}. Here, we discuss new perspectives that the coupling between SPs and nearby quantum emitters (QEs) open in the context of nanoscale quantum optics\cite{Tame:2013} 
and photo-emission tailoring\cite{Bozhevolnyi:2017b}.

Efficient and bright single-photon sources\cite{Andersen:2017}
constitute one of the enabling technologies for quantum
communication, computation, and information processing\cite{Pelton:2015,Lodahl:2015}, while being also indispensable for several configurations exploited within quantum sensing and imaging. In general, single-photon generation with high emission rates and fidelity is central to applications in quantum technologies, but its implementation is challenging due to the extremely weak interaction between single photons and individual QEs. 
A novel strategy to overcome this limitation is the utilization of strongly confined plasmonic, dielectric or hybrid electromagnetic modes that are efficiently interfaced with conventional optical waveguides from the infrared to the visible. The potential of this approach has been partially revealed in recent studies, while still not exhibiting it to the full extent due to the absence of a well established reliable technology. We anticipate that efficient interfaces between photons, SPs and QEs, forming the basis for quantum networks and enabling optical nonlinearities at the single-photon level\cite{Giannini:2011}, will mature and become a reality for practical use within a few years.

\section{Experimental Advances: Nano\-fabrication and Characterization Challenges}

\begin{figure}
\begin{center}
\includegraphics[angle=270,width=0.95\columnwidth]{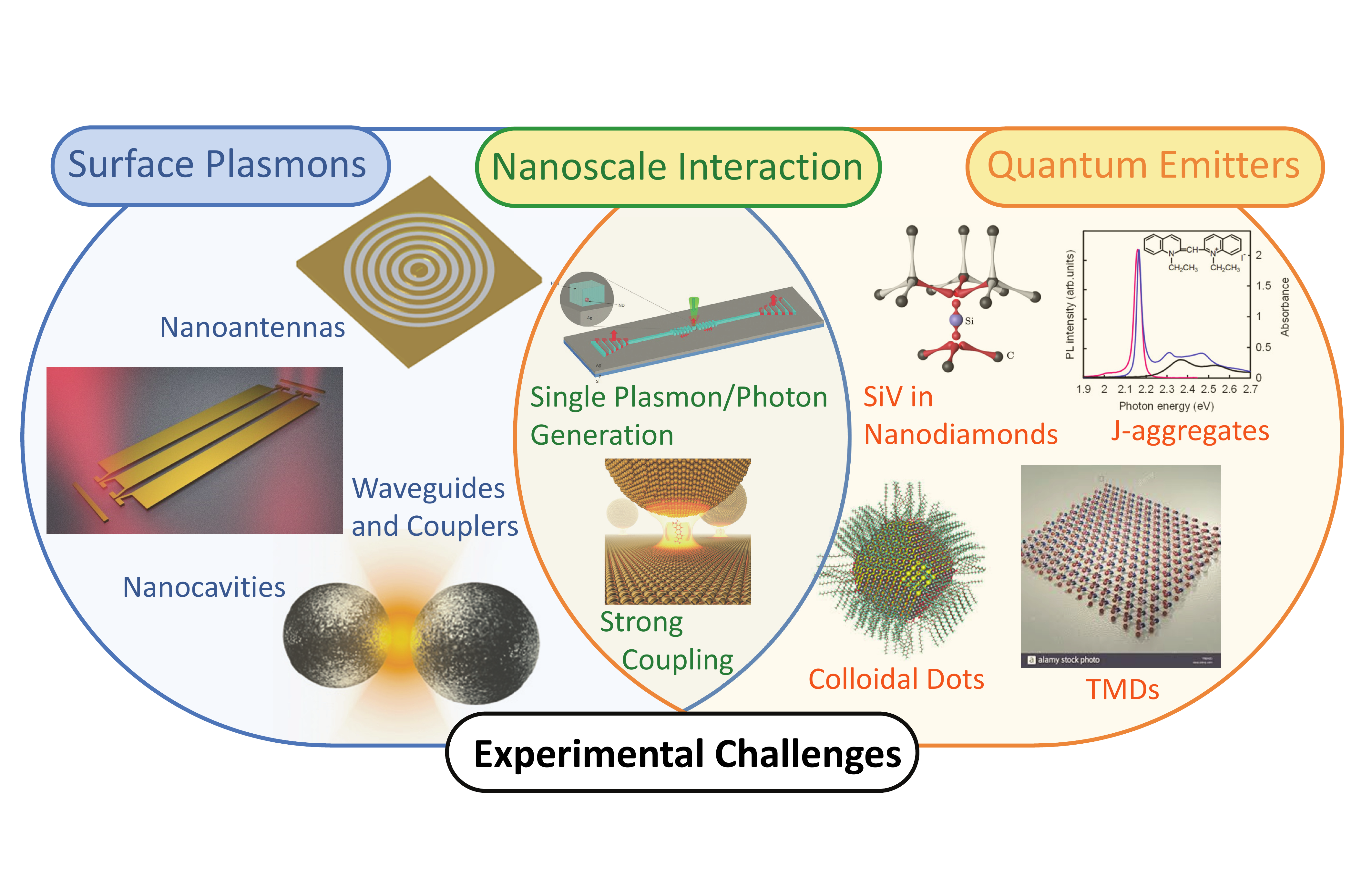}
\end{center}\linespread{1.3}
\caption{The left panel (blue region) encompasses different
nanophotonic devices, while the right one (orange) includes
various microscopic quantum light sources. The intersecting panel
(green) represents the main challenge towards the nano-scale
generation of non-classical light: the merging and hybridization
of photonic/plasmonic nanostructures and quantum light
sources.}\label{fig1}
\end{figure}

Reducing natural QE lifetimes is imperative for the realization of quantum photon sources with high repetition rates. This is generally done by placing them in a suitable photonic environment with an increased electromagnetic local density-of-states\cite{Pelton:2015,Lodahl:2015,Bozhevolnyi:2016}.
Clear objectives for future research will be making the best use
of the radiative and non-radiative Purcell enhancement accessible
in plasmonic cavities\cite{Chikkaraddy:2016} and antennae by transferring it into free space\cite{Curto:2010} or into photonic waveguides\cite{Chang:2006,Akimov:2007,Kumar:2013,Bermudez-Urena:2015} so as to significantly exceed the state-of-the-art quantum photon fluxes\cite{Hoang:2016}. A wide range of optical nano-devices (cavities, antennae and waveguides) are sketched in the left panel of Figure~\ref{fig1}. Demonstration of nonlinear single-photon operation is an outstanding challenge\cite{Sipahigil:2016}, and the incorporation and combination of these plasmonic/photonic structures would represent a major step towards the development of quantum optical networks with high repetition rates.

The right panel of Figure~\ref{fig1} displays some of the items in the current toolbox of quantum light sources: dye molecules, quantum dots, vacancy centers, and transition metal dichalcogenide (TMD) monolayers.  The fabrication of these QEs requires different technologies and their performance is subjected to different limitations. Due to their large stability, versatility and room-temperature operation, we anticipate further integration of solid-state single-photon QEs\cite{Aharonovich:2016}, such as colour
centres (NV, SiV, GeV and Sn) in nanodiamonds, defect centers in
2D materials, and nonlinear microcrystals in nanostructured environments. Besides single-photon emitters, entangled photon-pair production in nonlinear crystals is a key resource for optical quantum technologies. Although there is an increasing effort to realize integrated architectures on optical chips\cite{Grassani:2015}, there is so far no demonstration of photon-pair emission from nano- or micro-sized plasmon-enhanced systems.

Many quantum optical applications would benefit from having access
to regular trains of single photons with sufficiently high
repetition rates, ultimately delivering a flux of $\sim 10^9$
photons/s. Ideally, this flux should be seamlessly coupled to a
waveguide, so as to enable waveguide-integrated single-photon
sources with large Purcell enhancement ($\gg 10$) and efficient QE
coupling to the waveguide mode ($> 80\%$). In passing, we note that there can be a practical limit to how large Purcell factor one should strive for; if eventually too large, this could in principle promote the emitting of two photons per pulse, which would naturally be detrimental to single-photon applications\cite{Somaschi:2016}. For a detailed analysis of plasmonic environments, including key parameters which constitute a good efficient single photon source, we refer to Refs.~\cite{Bozhevolnyi:2016,Bozhevolnyi:2017b}. We envisage a
practical demand for QE integration into a hybrid nanophotonic
circuitry, e.g., pumped by an integrated laser diode with a reasonable overall efficiency from the photon emission to its detection ($>50\%$). Thus, plasmonic circuitry and metallo-dielectric waveguide configurations\cite{Ebbesen:2008,Smith:2015} (examples are shown in the left panel of Figure 1) are excellent candidates for ultra-compact single-photon optical sources and switches (possibly at cryogenic temperatures). Theoretical predictions indicate that the large coupling strengths attainable in plasmonic cavities allow not only the survival but also the enhancement of quantum nonlinearities in mesoscopic QE ensembles\cite{Saez-Blazquez:2017}, well beyond the single-emitter level.

A scientific and technological breakthrough that should
be targeted in coming years is the modularization of nanoscale
quantum optics circuitry, removing thereby severe roadblocks for
accepting quantum light sources for commercial use in emerging
photonic technologies. Truly non-classical light sources will need
to operate stand-alone and at room temperature, with the typical
brightness approaching the regime of nanowatts, i.e., rendering
detection with sensitive analog detectors possible. This should be
helped by technology platforms that allow merging of passive
dielectric and metallic structures with single QEs in a scalable
way\cite{Siampour:2017a,Siampour:2017b,Siampour:2018}. Further developments along this road would enable large flexibility in terms of possible quantum photonic networks without specific material constraints, thereby also paving the way for complementarity with existing commercial technology. As an example, using such a platform for nonlinear single-photon processes that require cryogenic operation (such as switches and logic gates) will imply that only a very small payload has to be eventually cooled. The central panel of Figure~\ref{fig1} highlights two promising avenues enabled by the marriage between QEs and the strong near-fields of plasmonic nanostructures: \emph{(i)} fast and bright ultra-compact light sources of single and/or entangled photons by spontaneous parametric down-conversion in nonlinear
microcrystals\cite{Benson:2011}, and \emph{(ii)} the realization of strong and ultra-strong coupling polaritonic phenomena\cite{Ramezani:2017}.

\section{Theoretical Advances: Fundamental Implications and Design Opportunities}

\begin{figure}
\begin{center}
\includegraphics[angle=270,width=0.95\columnwidth]{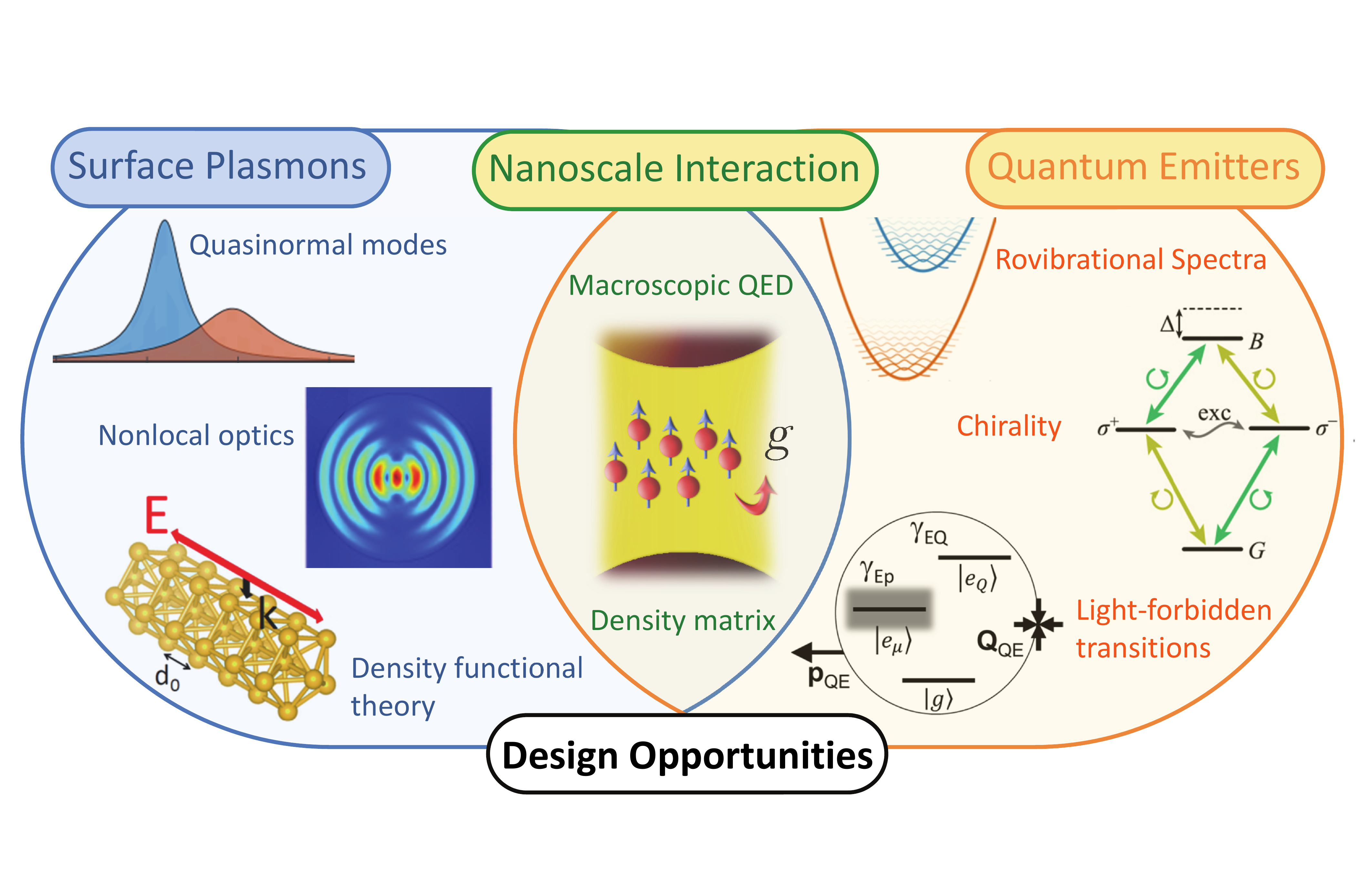}
\end{center}\linespread{1.3}
\caption{The left panel (blue region) shows different levels of
theoretical description of plasmonic devices beyond standard numerical solutions to Maxwell Equations. The right panel (orange) ranges various key aspects of microscopic light sources beyond their
two-level-system modeling. The intersecting panel (green)
illustrates the quantum optical design opportunities that resides
in the internal complexity of SPs and QEs, accessible through
their the nano-scale coupling.}\label{fig2}
\end{figure}

One fascinating aspect of light-matter interactions in
nanophotonic devices is the potential coexistence and exhibition
of quantum phenomena in both the light and
matter\cite{Bozhevolnyi:2017b}: the quantum optics associated
with light fields in proximity from metallic
nanostructures\cite{Marquier:2017}, and the quantum aspects of SPs
when being spatially confined to volumes approaching atomic-scale
dimensions\cite{Zhu:2016}. As illustrated in the left panel of
Figure~\ref{fig2}, the quest for harvesting the full potential of plasmonics for the generation of non-classical light calls for theoretical methods ranging from \emph{ab-initio} density-functional theory\cite{Varas:2016} and quantum corrected electrodynamics\cite{Christensen:2017,Yan:2015} to more semi-classical accounts with nonlocal corrections  of the classical electromagnetics, which is usually the starting point in state-of-the-art computational photonics. In general, \emph{ab-initio} descriptions seek to account for the microscopic electronic and atomistic degrees of freedom of the optical cavities\cite{Zuloaga:2009,Marinica:2015,Barbry:2015}. The quantum-corrected models invoke surface parameters, such as Feibelman parameters\cite{Christensen:2017} or artificial interface layers\cite{Yan:2015,Luo:2013}, to account for surface scattering associated with nonlocal response, quantum spill-out, or the relaxation of charge-transfer tunneling currents in gap structures\cite{Zhu:2016,Esteban:2012}. Finally, semi-classical descriptions essentially rely on hydrodynamic equations-of-motion (coupled with Maxwell equations) that account for the induced charge dynamics for a finite compressibility of the quantum electron gas\cite{Mortensen:2014}, and possibly also quantum spill-out included in a self-consistent manner\cite{Toscano:2015}.

At the emitter level, we emphasize in the right panel of Figure~\ref{fig2} different models for microscopic light sources beyond the archetypal two-level system for QEs. The subwavelength character of SP resonances give access to features in the QE population dynamics that remain hidden to propagating fields, effectively increasing its complexity and versatility as a light source. Thus, the large plasmonic spatial gradients yield light-matter interactions beyond the dipole approximation\cite{Benz:2016}, opening the way towards harvesting and operating with light-forbidden
exciton transitions\cite{Rivera:2016}. Similarly, exploiting the SP
sensitivity to both light and emitter polarizations, nonreciprocal
chiral coupling can be achieved among not only distant but also
neighboring QEs\cite{Lodahl:2017}. The ability of SPs to bridge
distinct length and energy scales also manifest in their
interaction with matter. It has been recently shown that, despite
lying at much lower energies than electronic transitions,
rovibrational molecular excitations can also be interact strongly
with SPs\cite{Memmi:2017}. These findings open new avenues not only
for photonic technology, but also for chemical applications\cite{Hutchison:2012}. Finally, the dispersive and propagating nature of excitons in 2D materials also needs to be tackled accurately in the description of polaritonic physics in bidimensional optoelectronic platforms, such as graphene or TMD monolayers\cite{Cuadra:2018}.

As illustrated in the central panel of Figure~\ref{fig2}, a wide range of design opportunities for quantum optical devices emerge by
exploiting the complexity of highly confined electromagnetic
fields and microscopic light sources. Accurate accounts of Purcell
enhancement and the further electromagnetic field quantization are
central for the description of nanoscale light-matter coupling. In
general, this must be done within the framework of macroscopic
quantum electrodynamics, which accounts for the open and lossy nature of both SPs and QEs. In order to reduce the complexity of the master
equation governing the density matrix of these systems, the method
of quasi-normal mode expansion\cite{Kristensen:2014,Lalanne:2018} has proven to be a convenient and insightful tool. The extension to also nonlocal hydrodynamic models was done only recently\cite{Dezfouli:2017}. An effort in quantizing these modes, which would indeed be a major milestone in quantum optics in general, would then allow a proper analysis and understanding of also multi-photon regimes, in the context of both single photon and entangled photon sources. For the multi-photon dynamics, we note the possibility for turning to so-called pseudomodes; while perhaps less intuitive, especially in the continuum regime with overlapping modes, they might be advantageous in cases where common semiclassical approaches fail\cite{Hughes:2018}. In certain geometries, quasi-analytical solutions to Maxwell Equations, such as those obtained by transformation optics\cite{Li:2016}, can also be employed to calculate key parameters such as the SP-QE coupling strengths or the SP lifetimes.

\section{Outlook}
In this article, we have reviewed some of the experimental challenges and design opportunities brought by nanophotonics science and technology when pushed into the quantum optical regime. On the one hand, despite recent progress, significant advances in current fabrication and characterization techniques will be required in order to control precisely the strong light-matter coupling taking place in hybrid systems comprising plasmons and quantum emitters, as well as to extract and utilize the quantum statistics of the light emerging from them. Beyond the continued use of noble metals, we anticipating new material avenues with more detailed materials processing\cite{McPeak:2015}, while also harvesting from alternatives\cite{Knight:2014,Bonanni:2011,Biggins:2018} as well as more dilute, tunable semiconductor and 2D-materials systems\cite{Liu:2017,Grigorenko:2012}. On the other hand, the deeply subwavelength nature of surface plasmons gives access to material features, both in their supporting metal nanostructures and the nearby microscopic light emitters, that remain hidden to propagating light. In the coming years, much theoretical efforts will focus on the exploitation, for photonic applications, of the novel material degrees of freedom made available in nanoscale quantum optical devices.    

\section*{Funding Sources}

A.~I.~F.-D. acknowledges funding
from EU Seventh Framework Programme [FP7-PEOPLE-2013-CIG-63099],
and the Spanish MINECO [FIS2015-64951-R and MDM-2014-0377]. S.~I.~B. acknowledges the European Research Council (grant 341054,
PLAQNAP). N.~A.~M. is a VILLUM Investigator supported by VILLUM FONDEN (grant 16498). Center for Nano Optics is financially supported by the University of Southern Denmark (SDU 2020 funding).

\newpage
%\bibliography{references}
\providecommand{\latin}[1]{#1}
\makeatletter
\providecommand{\doi}
  {\begingroup\let\do\@makeother\dospecials
  \catcode`\{=1 \catcode`\}=2\doi@aux}
\providecommand{\doi@aux}[1]{\endgroup\texttt{#1}}
\makeatother
\providecommand*\mcitethebibliography{\thebibliography}
\csname @ifundefined\endcsname{endmcitethebibliography}
  {\let\endmcitethebibliography\endthebibliography}{}

\newpage

\end{document}